# The current-induced spin-orbit torque and field-free switching from Mo-based magnetic heterostructures

Tian-Yue Chen[1], Hsin-I Chan[1], Wei-Bang Liao[1], and Chi-Feng Pai[1,2*]

[1]*Department of Materials Science and Engineering, National Taiwan University, Taipei 10617, Taiwan*

[2]*Center of Atomic Initiative for New Materials, National Taiwan University, Taipei 10617, Taiwan*

Magnetic heterostructure Mo/CoFeB/MgO has strong perpendicular magnetic anisotropy and thermal stability. Through current-induced hysteresis loop shift measurements, we show that the dampinglike spin-orbit torque (SOT) efficiency of Mo/CoFeB/MgO heterostructures is $\xi_{DL} \approx -0.003 \pm 0.001$ and fairly independent of the annealing temperature from 300°C to 400°C. Though $|\xi_{DL}|$ is small while compare to those from Ta or W-based heterostructures, reversible current-induced SOT switching of a thermally-stable Mo/CoFeB/MgO heterostruture can still be achieved. Furthermore, we observe field-free current-induced switching from a Mo/CoFeB/MgO structure with the Mo layer being wedge-deposited. Our results indicate that even for a weak spin-orbit interaction 4d transition metal such as Mo, it is still possible to generate sufficient spin current for conventional SOT switching and to realize field-free current-induced switching by structural engineering.

[*] Email: cfpai@ntu.edu.tw

# I. INTRODUCTION

In the development of contemporary spin-transfer torque magnetoresistive random access memory (STT-MRAM), perpendicular magnetic tunnel junction (p-MTJ) is an essential device element, which has a better scalability and thermal stability while compare to its in-plane magnetized predecessor [1]. Among many magnetic heterostructures that can give rise to the perpendicular magnetic anisotropy (PMA), Ta/CoFeB/MgO trilayer is one of the most studied and widely employed structures due to its simplicity and the compatibility with high tunneling magnetoresistance MgO junctions [2,3]. More importantly, not only for p-MTJ and current-perpendicular-to-plane STT-MRAM applications, Ta/CoFeB/MgO heterostructure has also been shown to have significant spin-orbit torque (SOT) efficiency, which can be utilized for SOT-induced magnetic switching [4] and microwave generation [5] with the current-in-plane scheme. The heavy transition metal (HM)/CoFeB/MgO PMA heterostructures, with HM = Ta, W [4,6,7], or Pt [8] (materials with large spin-Hall effects [9,10]), have become standard heterostructures for SOT studies. Characterizing the SOT efficiencies from various HM/CoFeB/MgO structures therefore can provide valuable information in building more energy-efficient SOT-MRAM devices.

On the other hand, enhancing the thermal stability of the magnetic memory element, which is the ferromagnetic CoFeB layer in HM/CoFeB/MgO heterostructures, plays a critical role in achieving stable STT-MRAM or SOT-MRAM devices. Besides Ta, several other HM materials have also been demonstrated to be suitable buffer layers for enhancing the PMA in HM/CoFeB/MgO structures. For instance, thermally annealed Hf/CoFeB/MgO [11] and Mo/CoFeB/MgO [12,13] have both been shown to gain greater PMA energy densities while compare to that of Ta/CoFeB/MgO. Surprisingly, Mo/CoFeB/MgO heterostructure can survive thermal annealing up to 450˚C, which makes it an attractive candidate for replacing Ta as the standard p-MTJ buffer layer structure that can be compatible with modern CMOS processing [12-15]. Although Mo has been introduced as a PMA-enhancing insertion layer for a SOT study published by Wu *et. al.* [16], its own SOT properties from a Mo/CoFeB/MgO structure without other HM underlayers have yet to be reported.

In this work, we perform characterizations on the magnetic properties and the dampinglike SOT (DL-SOT) efficiencies from Mo/CoFeB/MgO heterostructures with different annealing temperatures. We show that the DL-SOT efficiency originates from a thermally-annealed Mo/CoFeB/MgO heterostructure with PMA is $|\xi_{DL}| = 0.003 \pm 0.001$ with the sign being negative, which is consistent with previous reports on the spin Hall angle of Mo [17-19]. The DL-SOT efficiency of Mo/CoFeB/MgO is also found to be independent of the annealing temperature within the range from 300˚C to 400˚C. By assuming the observed SOT originates from the spin Hall effect (SHE) of Mo and a 100% spin transparency at the Mo/CoFeB interface, we estimate the lower bound of Mo spin Hall conductivity to be $|\sigma_{SH}| \approx 35(\hbar/e)\Omega^{-1}\text{cm}^{-1}$. Though $|\xi_{DL}| \approx 0.003$ of Mo-based structure is much smaller than those from Ta or W-based magnetic heterostructures ($|\xi_{DL}| \geq 0.10$), we can still observe reversible current-induced SOT switching in micron-sized Mo/CoFeB/MgO devices when an in-plane field is applied. More importantly, we show that if the Mo layer is wedge deposited, then an out-of-plane current-induced effective field will emerge even in the absence of in-plane bias field. Pure current-induced magnetization switching can therefore be realized in Mo(wedge)/CoFeB/MgO heterostructures. Our results suggest that although the strength of current-induced SOT is small in Mo/CoFeB/MgO heterostructures, it is not entirely negligible. In addition, by controlling the deposition method, it is possible to achieve field-free current-induced magnetization switching in HM/CoFeB/MgO PMA structures even if the HM is a 4d transition metal with weak spin-orbit interaction.

## II. MAGNETIC PROPERTIES CHARACTERIZATION

To characterize the magnetic properties of Mo/CoFeB/MgO heterostructures, we prepare a series of Mo(4)/Co$_{40}$Fe$_{40}$B$_{20}$($t_{\text{CoFeB}}$)/MgO(2)/Ta(2) on thermally-oxidized silicon substrate by high-vacuum magnetron sputtering. Numbers in the parenthesis represent nominal sputtered thickness of each layer and the Ta(2) serves as a capping layer to prevent oxidation. The base pressure of the deposition system is $\sim 3\times10^{-8}$ Torr and the Ar working pressures for sputtering metallic and oxide layers are set to be 3 mTorr and 10 mTorr, respectively. After depositions, the films are annealed in vacuum at 300˚C for 1 hour to induce PMA. A representative out-of-plane hysteresis loop obtained by magneto-optical Kerr effect (MOKE) from a Mo(4)/Co$_{40}$Fe$_{40}$B$_{20}$(1.4)/MgO(2)/Ta(2) film annealed at 300˚C is shown in Fig. 1(a), from which the PMA of the film is verified. We further characterize saturation magnetization $M_s$ and effective anisotropy energy density $K_{\text{eff}}$ of the annealed Mo/CoFeB/MgO heterostructures by vibrating sample magnetometer (VSM), as shown in Fig. 1(b) and (c). The effective saturation magnetization is found to be $M_s = 1200 \, \text{emu/cm}^3$ with a magnetic dead layer of $t_{dead} = 0.8 \, \text{nm}$. The large magnetic dead layer might originate from the intermixing of CoFeB and buffer layer at the Mo/CoFeB interface [20] or the partial oxidation or boron segregation at the CoFeB/MgO interface [21,22]. The $K_{eff} \cdot t_{CoFeB}^{eff}$ vs. $t_{CoFeB}^{eff}$ ($= t_{CoFeB} - t_{dead}$) plot also indicates that PMA exists when $0.4 \, \text{nm} \leq t_{CoFeB}^{eff} \leq 0.9 \, \text{nm}$ ($1.2 \, \text{nm} \leq t_{CoFeB} \leq 1.7 \, \text{nm}$), though the maximum $K_{eff} \cdot t_{CoFeB}^{eff}$ value of 0.03 erg/cm$^2$ is smaller than that reported by Liu *et. al.* [12] due to the existence of magnetic dead layer. Next, we fix the nominal CoFeB thickness at $t_{CoFeB} = 1.4 \, \text{nm}$ and study the dependence of Mo(4)/CoFeB(1.4)/MgO(2) PMA on annealing temperature. As shown in Fig. 1(d), PMA of Mo(4)/CoFeB(1.4)/MgO(2) exists for annealing temperature $T_a$ ranges from 280˚C to 420˚C and the out-of-plane coercive field peaks at around $T_a \approx 300$˚C. Therefore, we confirm that Mo/CoFeB/MgO heterostructures can survive thermal-annealing with $T_a$ greater than 400˚C, which is consistent with previous reports.

### III. DL-SOT EFFICIENCY CHARACTERIZATION

After verifying the existence of PMA in Mo/CoFeB/MgO structures over a wide range of annealing temperature, we pattern a Mo(4)/CoFeB(1.4)/MgO(2) film ($T_a$ = 300˚C) into micron-sized Hall-bar devices through photolithography. The nominal width of the Hall-bar is $w = 5$ μm. In order to estimate the SOT efficiency from those devices, we utilize current-induced hysteresis loop shift measurement to quantify the out-of-plane effective field $H_z^{\text{eff}}$, which is a manifestation of the DL-SOT acting on the chiral domain wall moments of CoFeB layer [8,23-27]. As schematically shown in Fig. 2(a), the Hall-bar device is subject to a static in-plane bias field $H_x$ to overcome the effective field $H_{\text{DMI}}$ originates from Dzyaloshinski-Moriya interaction (DMI). While increasing the magnitude of $H_x$, the DMI-caused Néel-type chiral domain wall moments in the CoFeB layer will gradually align with respect to $H_x$ and further facilitate DL-SOT-driven domain wall motion [28-30]. Once $H_{\text{DMI}}$ is overcome, the full strength of DL-SOT can be detected by the shift of out-of-plane hysteresis loops, which are recorded through anomalous Hall effect (AHE) from the Hall-bar device. Representative loop shift results are shown in Fig. 2(b). A current-induced effective field $H_z^{\text{eff}}$, which is proportional to the strength of DL-SOT, can be observed when $H_x = 800\,\text{Oe}$ is applied with a constant supply of DC current $I_{dc}$. The asymmetry of $\pm I_{dc}$ shown in Fig. 2 (b) is to avoid measuring AHE loops with $I_{dc} = 0\,\text{mA}$ while stepping down the applied current by 0.4 mA. As shown in Fig. 2(c), the ratio of $H_z^{\text{eff}} / I_{dc}$ saturates at around ±1 Oe/mA when the in-plane bias field $H_x \approx \pm 200\,\text{Oe}$. This suggests that $|H_{\text{DMI}}| \approx 200\,\text{Oe}$ and the DMI constant $|D| = \mu_0 M_s \delta |H_{\text{DMI}}| \approx 0.35\,\text{mJ/m}^2$ [31], where the domain wall width $\delta$ is estimated by $\delta = \sqrt{A/K_{\text{eff}}}$ and the exchange stiffness constant $A$ is assumed to be ~ $10^{-11}$ J/m. The magnitude of DMI for Mo/CoFeB is therefore smaller than the reported values of 0.6-2.9 mJ/m² for large DMI Pt/Co [8,32-35], which indicates that Mo has a weaker spin-orbit coupling while compare to Pt.

The saturated DL-SOT efficiency $\xi_{DL}$ of this Mo(4)/CoFeB(1.4)/MgO(2) Hall-bar device can be further estimated by [36]

$$\xi_{DL} = \frac{2e}{\hbar}\left(\frac{2}{\pi}\right)\mu_0 M_s t_{\text{CoFeB}}^{\text{eff}} w t_{\text{Mo}} \left(\frac{\rho_{\text{CoFeB}} t_{\text{Mo}} + \rho_{\text{Mo}} t_{\text{CoFeB}}}{\rho_{\text{CoFeB}} t_{\text{Mo}}}\right)\left(\frac{H_z^{\text{eff}}}{I_{dc}}\right), \quad (1)$$

where $\rho_{\text{CoFeB}} = 180\,\mu\Omega\text{-cm}$ and $\rho_{\text{Mo}} = 85\,\mu\Omega\text{-cm}$ are the resistivities of CoFeB layer and Mo layer, respectively. The resistivities are determined by measuring the resistance of Mo/CoFeB/MgO heterostructures with various Mo layer thicknesses. Using the VSM-determined $M_s = 1200\,\text{emu/cm}^3$ (S.I. unit $1.2\times 10^6$ A/m), $t_{\text{CoFeB}}^{\text{eff}} = 0.6\,\text{nm}$, and the loop-shift-determined $H_z^{\text{eff}}/I_{dc} \approx -1\,\text{Oe/mA}$, the DL-SOT efficiency is estimated to be $\xi_{DL} \approx -0.003$. This $|\xi_{DL}| \approx 0.003$ is the lower bound of the intrinsic spin Hall ratio of Mo due to (1) imperfect spin transmission at the HM/CoFeB interface [37,38] and (2) possible current shunting in the Hall-bar device structure [39,40], which typically leads to an underestimation of the DL-SOT efficiency. The lower bound of Mo spin Hall conductivity is therefore calculated to be $|\sigma_{\text{SH}}| \geq |\xi_{DL}/\rho_{\text{Mo}}| \approx 35(\hbar/e)\,\Omega^{-1}\text{cm}^{-1}$. More importantly, the estimated DL-SOT efficiency $|\xi_{DL}|$ of Mo(4)/CoFeB(1.4)/MgO(2) heterostructure is fairly independent of annealing temperature $300\,^\circ\text{C} \leq T_a \leq 400\,^\circ\text{C}$, as summarized in Fig. 2(d). The $T_a$-independent $|\xi_{DL}| = 0.003 \pm 0.001$ of Mo/CoFeB/MgO system suggests that Mo is a thermally-stable buffer layer for studying both PMA and SOT for a wide range of annealing temperature.

## IV. CURRENT-INDUCED DL-SOT SWITCHING

Although the DL-SOT efficiency from Mo-based magnetic heterostructures is two orders of magnitude smaller than those from Pt, Ta, and W-based magnetic heterostructures [4,6,41,42], it is still possible to realize current-induced switching. To demonstrate SOT-driven magnetization switching, we apply current pulses with pulse width $0.01\,\text{s} \leq t_{\text{pulse}} \leq 1\,\text{s}$. Current-induced switching results recorded from a 10 μm-wide, $300\,^\circ\text{C}$-annealed Mo(4)/CoFeB(1.4)/MgO(2) Hall-bar device with $t_{\text{pulse}} = 0.1\,\text{s}$ and critical switching current $|I_c| \approx 10\,\text{mA}$ are shown in Fig. 3(a) and (b). The opposite switching polarity with respect to the applied in-plane field $H_x = \pm 800\,\text{Oe}$ is consistent with the symmetry of SOT-driven dynamics [4,29,42]. Since SOT-driven magnetization switching is a thermally-activated process, $I_c$ should depend on the applied current pulse width and can be expressed as [43]

$$I_c = I_{c0}\left[1 - \frac{1}{\Delta}\ln\left(\frac{t_{\text{pulse}}}{\tau_0}\right)\right], \tag{2}$$

where $I_{c0}$ is the zero-thermal critical switching current, $\Delta \equiv U/k_B T$ is the thermal stability factor ($U$ is the energy barrier), and $1/\tau_0 \approx 1\,\text{GHz}$ ($\tau_0 \approx 1\,\text{ns}$) is the attempt rate for thermally-activated switching [44]. As shown in Fig. 3(c), linear trends can be found in the experimental $I_c$ vs. $\ln(t_{\text{pulse}}/\tau_0)$ plot. By performing linear fits of the switching data with Eqn. (2), we estimate $|I_{c0}| \approx 20\,\text{mA}$ and $\Delta \approx 41$ for this Mo(4)/CoFeB(1.4)/MgO(2) Hall-bar device. The zero-thermal critical switching current density is further calculated to be $|J_{c0}| \approx 4.2 \times 10^{11}$ A/m$^2$. By comparison, as shown in Fig. 3(d), the results from a W(4)/CoFeB(1.4)/MgO(2) control sample indicate a similar thermal stability of CoFeB but a much lower critical switching current density $|J_{c0}| \approx 7.6 \times 10^{10}$ A/m$^2$ due to the greater SHE from W [6,7].

## V. FIELD-FREE CURRENT-INDUCED SWITCHING

Beyond conventional SOT switching of perpendicular magnetization, which typically requires an in-plane bias field, there has been an increasing interest to realize current-induced switching in the absence of external field [45-50]. Though the origin of current-induced out-of-plane effective field $H_z^{eff}$ is still elusive, it has been shown that by introducing a wedge structure in the oxide layer of Ta/CoFeB/TaO$_x$ system, such effective field can be utilized to switch magnetization deterministically without applying any in-plane field [51,52]. Similar mysterious $H_z^{eff}$ has also been seen recently in a Pt/W/CoFeB/MgO heterostructure [53], which was attributed to the competing spin currents from the Pt and W buffer layers. Interestingly, the W layer in Ref. [53] is also oblique deposited. To unveil the main cause of wedge-structure-induced $H_z^{eff}$ and to understand if strong spin-orbit interaction materials are indeed necessary, we perform DL-SOT efficiency characterizations for two Mo-based samples: Ta(2)/MgO(2)/CoFeB(1.4)/Mo(10) and Ta(2)/MgO(2)/CoFeB(1.4)/Mo(wedge). Wedge deposition is done by turning off sample holder rotation during the sputtering of Mo layer, and the nominal thickness of Mo at the center of the film is ~7 nm. Both samples are annealed at 300 $^{\circ}$C for 1 hour to induce PMA and patterned into Hall-bar devices with $w = 5$ μm. The layer structure and thickness gradient of the wedge-deposited sample and its relative orientation to the applied current/field during measurement are shown in Fig. 4(a). Note that we prepared the film in reverse order (MgO/CoFeB/HM) to make sure MgO and CoFeB layers are flat with respect to the substrate and only the Mo layer is wedged. The magnetic dead layer thickness is determined to be $t_{dead} = 0.3$ nm by VSM measurements.

In Fig. 4(b), we summarize the measured $H_z^{eff}/I_{dc}$ for both uniform and wedge-deposited samples under different in-plane bias fields $H_x$ through current-induced loop shift measurements. Although the overall $H_x$ dependence of $H_z^{eff}/I_{dc}$ for both samples are similar, it is found that there exists a finite $H_z^{eff}/I_{dc} \approx 0.5$ Oe/mA at $H_x = 0$ Oe for the wedge deposited device. In contrast, $H_z^{eff}/I_{dc} \leq 0.2$ Oe/mA for the uniform case when the in-plane field is absent. Note that the measured $|H_z^{eff}/I_{dc}|$ becomes smaller when the applied $H_x \geq 400$ Oe. It is possible that if the applied $|H_x|$ is much greater than $|H_{DMI}|$, then the magnetic moments in the film become more single-domain-like and therefore the SOT-driven or SOT-assisted domain nucleation/propagation is not as efficient as the intermediate $H_x$ case. This slight decrease of DL-SOT efficiency at $|H_x| \gg |H_{DMI}|$ has also been seen in some recent works [54,55].

Due to the existence of a finite current-induced $H_z^{eff}$ at $H_x = 0$ Oe in the wedge-deposited Mo device, deterministic current-induced magnetization can be achieved, though Mo is a transition metal with weak spin-orbit interaction. As shown in Fig. 4(c), the representative AHE result indicates that the MgO(2)/CoFeB(1.4)/Mo(wedge) Hall-bar device has an overall Hall resistance difference of $\Delta R_H = 0.6\,\Omega$ for full magnetization switching. We observe a reversible and robust field-free current-induced magnetization switching of $\Delta R_H = 0.45\,\Omega$ from the very same wedge-Mo device with $I_{sw} \approx \pm 3.8\,\text{mA}$, as shown in Fig. 4(d). The difference of $\Delta R_H$ between field-induced and current-induced switching is due to the device area that cannot be switched by current, as indicated by the insets of Fig. 4(c) and (d). The field-free switching of the wedge Mo device is performed on a probe station without electromagnets, which prevents the influence of remnant fields from magnetic pole pieces. We observe no field-free current-induced magnetization switching in the MgO/CoFeB/Mo devices prepared by uniform deposition. Our results suggest that as long as the HM layer is deposited in an oblique way, there exists a non-negligible current-induced $H_z^{\text{eff}}$ that can be employed to realize field-free switching. The HM layer is not necessarily being materials with significant spin-orbit interaction or SHE, such as Ta, W, or Pt.

Although conventional current-induced torques from strong spin-orbit interactions might not play important roles for the observed field-free switching in Mo-based devices, the Oersted field might play a crucial role. It has been shown that current-induced Oersted field has a non-negligible z-component at the edges of Hall-bar sample [56], which will further create a longitudinal domain wall in some cases [57]. It has also been shown by simulations that an effective out-of-plane field will be created by the magnetic anisotropy gradient, which can be utilized to drive domain wall motion [58,59]. Therefore, we postulate that the current that we applied will first create a longitudinal domain wall in the sample, and then the magnetization switching will be achieved by lateral domain wall motion, which is driven by the transverse magnetic anisotropy gradient.

The oblique deposition or the glancing angle deposition (GLAD) of Mo layer might also favor a tilted columnar microstructure [60,61], which modifies the Mo crystal structure symmetry. It is therefore possible that an extra dampinglike SOT will be generated when applying current along a low-symmetry axis of Mo, which is similar to the phenomenon reported in a recent study on WTe$_2$/Py system [62]. We hope future works that involve detailed MOKE imaging and microstructure analysis can further elucidate the real microscopic origin of wedge-direction-dependence field-free switching.

## VI. CONCLUSION

To conclude, we show that Mo/CoFeB/MgO magnetic heterostructure has decent PMA, which is robust against thermal annealing within temperature range of $280\,^\circ\text{C} \leq T_a \leq 420\,^\circ\text{C}$. The DL-SOT efficiency in such system is estimated to be small but also fairly independent of annealing temperature ($300\,^\circ\text{C} \leq T_a \leq 400\,^\circ\text{C}$), $|\xi_{DL}| = 0.003 \pm 0.001$. The lower bound of spin Hall conductivity of Mo, $|\sigma_{SH}| \approx 35 (\hbar/e) \Omega^{-1}\text{cm}^{-1}$, is therefore small while compare to other strong SHE metals such as Pt, Ta, and W. Despite the weak spin-orbit interaction of 4d transition metal Mo, conventional current-induced SOT switching can still be achieved in micron-sized Mo/CoFeB/MgO Hall-bar devices with the aid of in-plane bias fields. Surprisingly, by depositing the Mo layer with an oblique angle, field-free current-induced magnetization switching can be realized. This discovery implies that strong spin-orbit interaction materials or spin Hall sources are not always necessary in achieving field-free current-induced switching, and "to wedge or not to wedge" during the thin film deposition process is a more critical factor.

**Acknowledgments**

This work is supported by the Ministry of Science and Technology of Taiwan (MOST) under grant No. MOST 105-2112-M-002-007-MY3 and by the Center of Atomic Initiative for New Materials (AI-Mat), National Taiwan University, Taipei, Taiwan from the Featured Areas Research Center Program within the framework of the Higher Education Sprout Project by the Ministry of Education (MOE) in Taiwan under grant No. NTU-107L9008. We thank Tsao-Chi Chuang for her support on AFM and VSM measurements. We also thank Prof. Chia-Ling Chien of Johns Hopkins University for fruitful discussions on the possible origins of field-free switching.

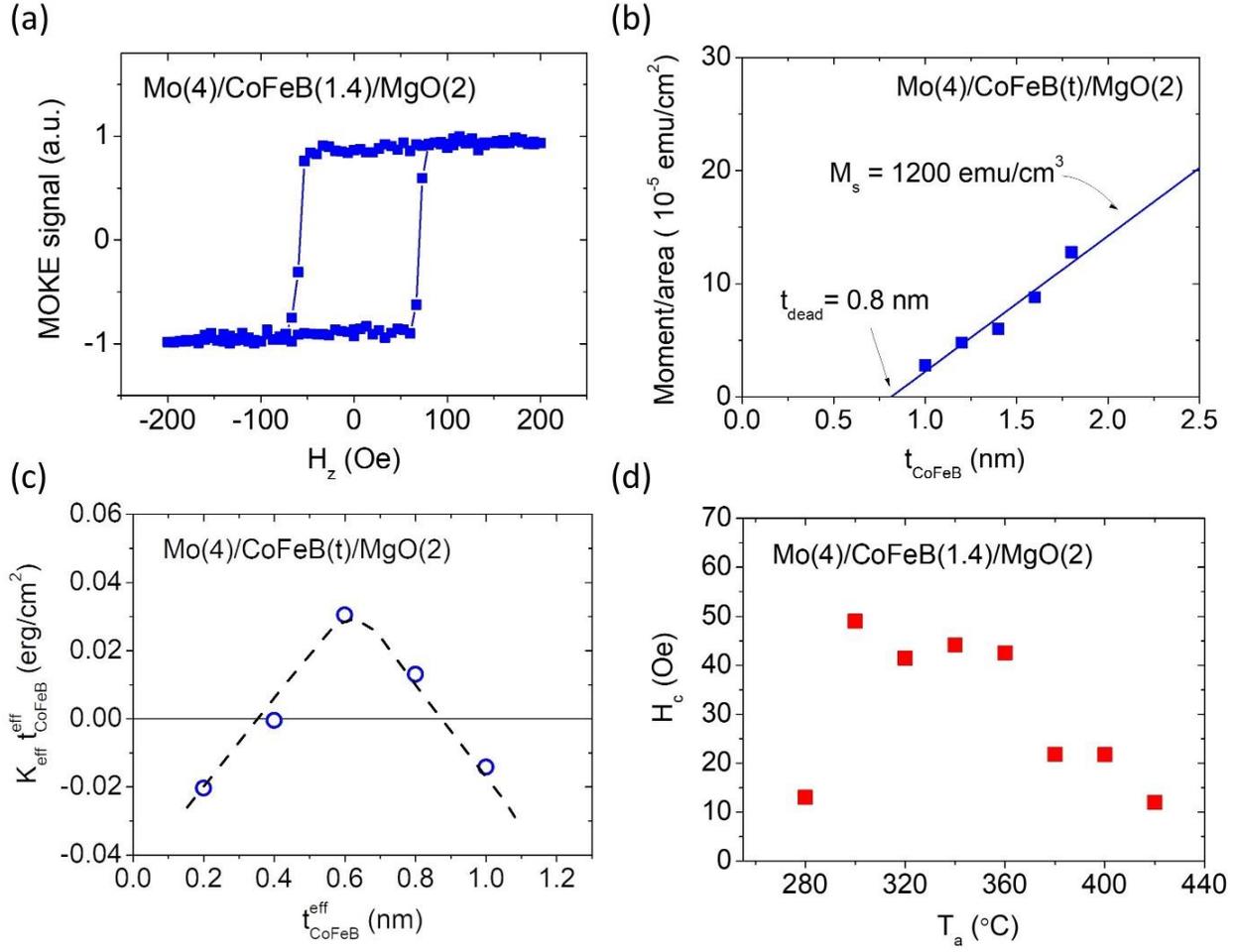

Figure 1. (a) Representative out-of-plane hysteresis loop of a Mo(4)/CoFeB(1.4)/MgO(2) film annealed at 300°C. (b) Magnetic moment per area as a function of CoFeB thickness $t_{CoFeB}$ for Mo(4)/CoFeB($t_{CoFeB}$)/MgO(2) films annealed at 300°C. The line represents linear fit to the data. (c) The product of effective magnetic anisotropy energy density and effective CoFeB thickness as a function of effective CoFeB thickness $t_{CoFeB}^{eff}$ for Mo(4)/CoFeB($t_{CoFeB}$)/MgO(2) films annealed at 300°C. The dashed line serves as guide to the eye. (d) Out-of-plane coercive field of Mo(4)/CoFeB(1.4)/MgO(2) samples as a function of annealing temperature $T_a$.

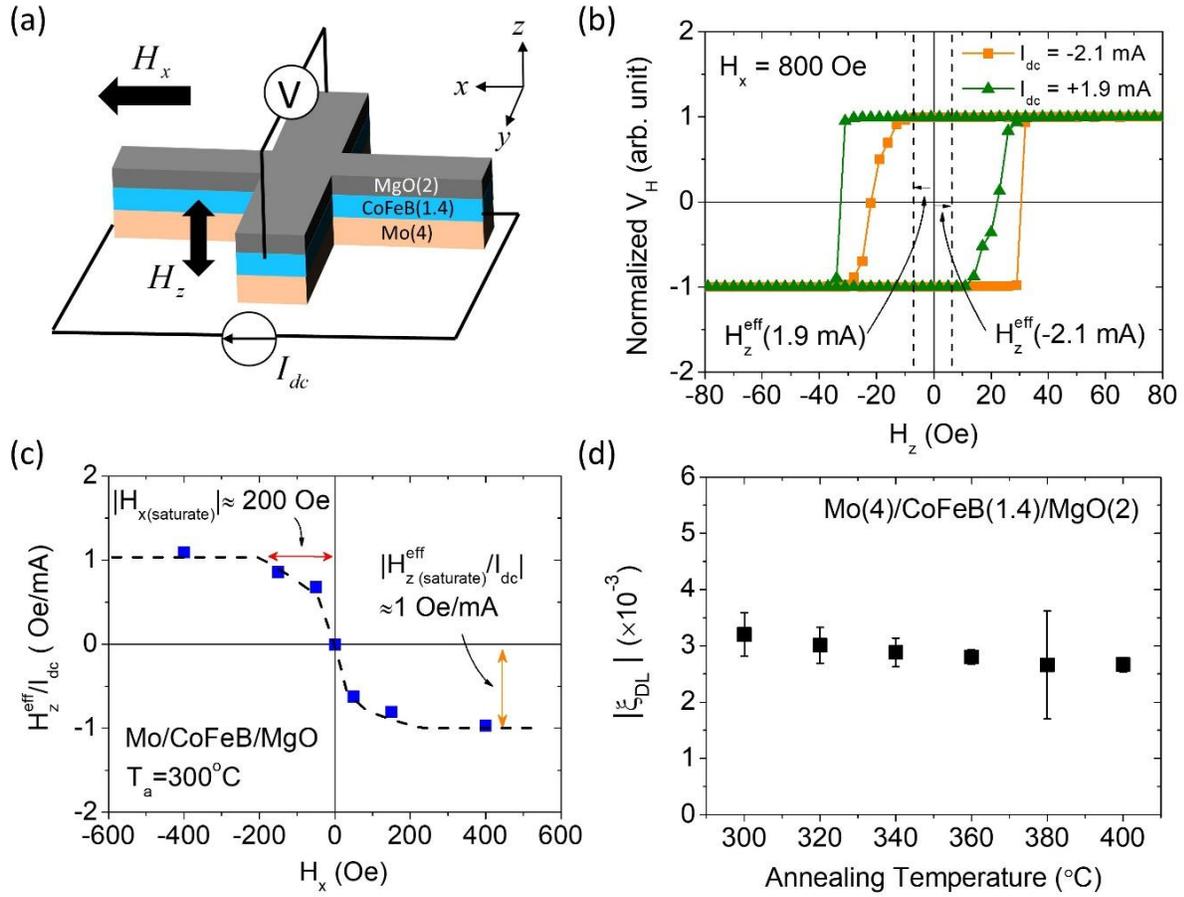

Figure 2. (a) Schematic illustration of patterned Mo/CoFeB/MgO Hall-bar structure and setup for current-induced hysteresis loop shift measurement. $H_x$ and $H_z$ stand for externally-applied in-plane and out-of-plane field, respectively. $I_{dc}$ represents the applied DC current. (b) Representative hysteresis loops obtained from a Mo(4)/CoFeB(1.4)/MgO(2) Hall-bar sample (annealed at 300˚C) through anomalous Hall effect with different $I_{dc}$ and in-plane bias field $H_x = 800\,\text{Oe}$. $H_z^{\text{eff}}$ represents the effective field stemming from DL-SOT. (c) $H_z^{\text{eff}}/I_{dc}$ as a function of in-plane bias field $H_x$. (d) The estimated magnitude of DL-SOT efficiency $|\xi_{DL}|$ for Mo(4)/CoFeB(1.4)/MgO(2) samples as a function of annealing temperature $T_a$.

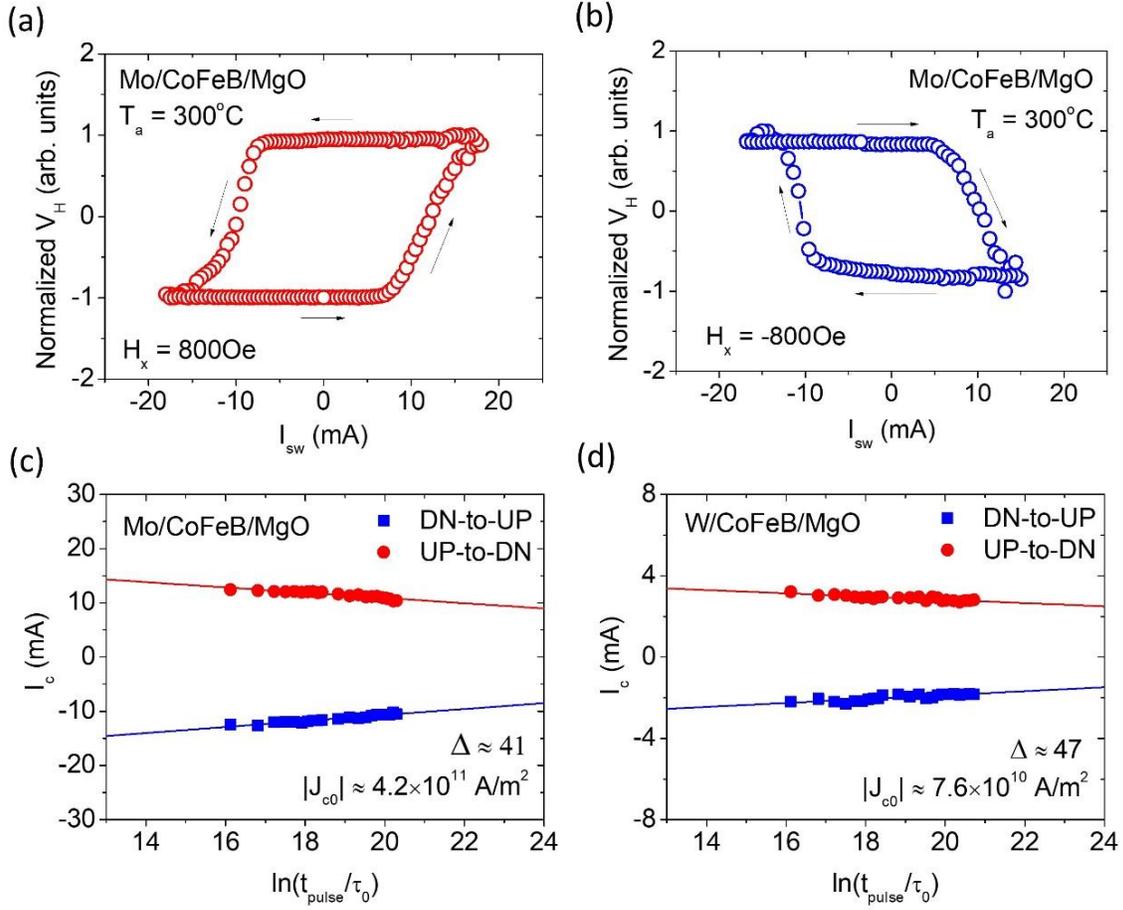

Figure 3. (a, b) Pulse current-induced switching of a 10 μm-wide Mo(4)/CoFeB(1.4)/MgO(2) Hall-bar sample annealed at 300˚C with in-plane bias field $H_x = \pm 800\,\text{Oe}$. The arrows indicate the switching directions. The applied current pulse width $t_{\text{pulse}}$ dependence of critical switching current $I_c$ for the (c) Mo(4)/CoFeB(1.4)/MgO(2) and (d) W(4)/CoFeB(1.4)/MgO(2) Hall-bar samples. Both are annealed at 300˚C. The solid lines represent linear fits to the experimental data.

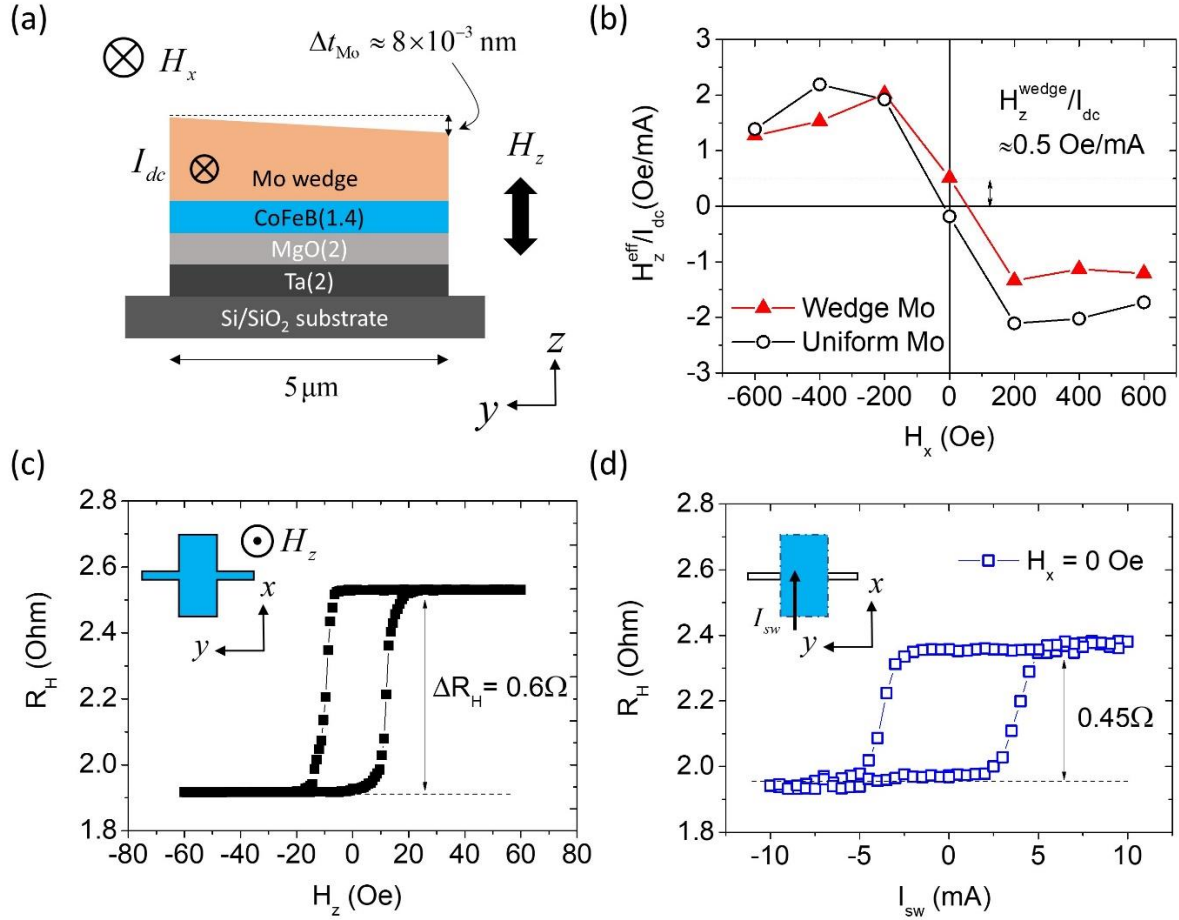

Figure 4. (a) Layer structure of the wedge-deposited Mo sample (not to scale). (b) $H_z^{eff}/I_{dc}$ as functions of in-plane bias field $H_x$ for uniform Mo (black circles) and wedge-deposited Mo (red triangles) Hall-bar samples. $H_z^{wedge}/I_{dc}$ represents the effective field per current that originates from the wedge structure. (c) Out-of-plane hysteresis loop and (d) field-free current-driven magnetization switching loop of a MgO(2)/CoFeB(1.4)/Mo(wedge) Hall-bar device obtained via anomalous Hall effect. The Hall-bar top-view illustrations indicate the area of magnetization being switched for each case.